\documentclass{article}
\usepackage{graphicx} 
\usepackage{authblk}
\usepackage{orcidlink}
\usepackage{amsmath}
\usepackage{booktabs}
\usepackage{tabularx}
\usepackage{hyperref}
\usepackage[labelfont=bf,font=small,textfont=it]{caption}

\title{ \bfseries Scheduler-Driven Job Atomization}
\author[1]{Michal Konopa\,\orcidlink{0000-0003-1694-1529}}
\author[1]{Jan Fesl\,\orcidlink{ 0000-0001-7192-4460}}
\author[1]{Ladislav Ber{\'a}nek\,\orcidlink{0000-0001-5004-0164}}

\affil[1]{Department of Data Science and Computer Systems, Faculty of Agriculture and Technology, University of South Bohemia, Czech Republic\\
\texttt{konopm05@fzt.jcu.cz}, \texttt{jfesl@fzt.jcu.cz}, \texttt{beranek@fzt.jcu.cz}}

\date{September 2025}

\begin{document}
\maketitle

\begin{abstract}
\noindent Modern GPU clusters, particularly those built on NVIDIA’s Multi-Instance GPU (MIG) architecture, often suffer from inefficiencies because jobs are treated as rigid, indivisible blocks that occupy a fixed slice until completion. The reliance on static peak memory estimates exacerbates fragmentation, underutilization, and job rejections.

We propose Scheduler-Driven Job Atomization (SJA), a new paradigm that establishes a bidirectional interaction between scheduler and jobs. In SJA, the scheduler advertises available execution gaps, and jobs respond by signaling interest if they can potentially generate a subjob that fits the offered time–capacity window. The scheduler may collect multiple signals for the same slot and, based on its allocation policy (e.g., fairness, efficiency, or SLA priorities), selects which job is granted the slot. Only then does the chosen job materialize a safe, self-contained subjob tailored to that opportunity.

Unlike migration or preemption, SJA proactively shapes workloads before execution, thereby avoiding costly state transfers and unpredictable interruptions. It aims to increase GPU utilization, reduce wait times, and minimize migration overhead by aligning jobs with opportunities in real time, ensuring that each admitted subjob is correct by construction.

This paper is presented as a concept paper: it introduces the paradigm, defines its building blocks, and outlines future research directions, rather than offering a full experimental evaluation.
\end{abstract}


\section{Introduction}
This paper sets out the conceptual foundations of Scheduler-Driven Job Atomization (SJA), focusing on defining the paradigm and motivating its relevance, while detailed evaluation is left for future studies.

The growing demand for GPU-accelerated computing in domains such as scientific research, machine learning, and Agriculture 4.0 has introduced new challenges for workload scheduling. NVIDIA’s \emph{Multi-Instance GPU (MIG)} technology, first released with the Ampere architecture, enables hardware-isolated partitions that allow multiple jobs to run concurrently on a single device. A GPU such as the A100 can be divided into up to seven independent slices, each with dedicated streaming multiprocessors, cache, and memory capacity~\cite{NVIDIA2025MIG}. From the user’s perspective, each slice operates as a smaller standalone GPU with guaranteed performance isolation.

This design improves consolidation and enables more fine-grained allocation than whole-GPU scheduling. However, MIG slices are typically configured statically, and their fixed sizes do not always align with the temporal and memory requirements of incoming workloads. As a result, fragmentation persists: some slices remain idle or underutilized even while other jobs wait in the queue.

Addressing these inefficiencies requires new scheduling approaches capable of adaptively exploiting fragmented capacity. Most current schedulers, however, still treat jobs as rigid and indivisible: each job is defined by a user-declared peak memory requirement and an estimated runtime, and once admitted, it occupies its assigned slice until completion. This rigidity results in idle capacity, wasted resources, and higher rejection rates for incoming workloads. By reconceptualizing jobs not as monolithic blocks but as adaptive, interactive entities, schedulers could more effectively align demand with fragmented capacity and improve overall system efficiency.

To this end, we propose SJA. In this paradigm, the scheduler advertises \emph{available execution gaps}, and jobs respond by signaling interest if they can safely generate a subjob for that slot. The scheduler may collect multiple signals for the same gap and, according to its policy (e.g., efficiency, fairness, or SLA priorities), select which job is granted the opportunity. Only then does the chosen job materialize a self-contained subjob tailored to the allocated window. This two-phase negotiation transforms scheduling from a passive allocation into an interactive protocol that exploits fragmentation without requiring costly mid-run reallocation.

The contribution of this paper is threefold: (i) to define the paradigm of SJA, (ii) to clarify its key building blocks (subjobs, temporal and functional memory profiles), and (iii) to articulate how bidirectional communication between scheduler and jobs can transform fragmented GPU capacity into safe and efficient execution opportunities. The core principles underlying SJA are summarized in Section~\ref{sec:core-principles}.

Unlike existing strategies such as migration, preemption, or moldable scheduling, SJA establishes a fundamentally new bidirectional model in which schedulers 
advertise fragmented opportunities, jobs signal interest, and only selected jobs generate subjobs that fit by construction. A dedicated discussion of how SJA 
relates to migration is provided in Section~\ref{sec:relation-migration}.

The scope of this work is deliberately conceptual: we assume a MIG-enabled GPU cluster, introduce Temporal Resource Profiles (TRPs) as a general mechanism for probabilistic resource prediction, and focus on Functional Memory Profiles (FMPs) as the primary instance, while leaving performance evaluation to future simulation and prototyping studies.

\section{Related Work and Novelty}
Prior efforts to improve GPU utilization span several directions. A comprehensive survey of GPU scheduling for deep learning workloads highlights diverse challenges and existing techniques, including resource sharing, fairness, and fragmentation~\cite{Ye2024CSUR}. While the focus is on DL applications, these findings confirm the central role of scheduling in multi-tenant GPU datacenters and motivate the search for more flexible paradigms.

(i) \emph{Malleable and moldable scheduling.} 
Malleable and moldable models allow a job’s resource footprint to change across runs—or, in the malleable case, even at runtime—so the scheduler can select a “shape” (e.g., number or size of GPU slices) that better fits available capacity. In the context of NVIDIA MIG, Tan et al.\ introduced MIG-Serving, framing it as a reconfigurable machine scheduling problem where jobs may be reassigned to different slice shapes~\cite{Tan2021MIGServing}. Villarrubia et al.\ extended this line of work by developing moldable scheduling heuristics explicitly tailored to MIG-enabled environments, where the shape is fixed prior to execution but chosen from multiple alternatives~\cite{Villarrubia2025JPDC}. Both studies show that adapting job shapes can mitigate fragmentation and improve overall throughput, yet once a job begins execution, it remains monolithic and cannot be decomposed into finer-grained subjobs.

(ii) \emph{Divisible-load theory (DLT).} 
DLT studies how large computations can be partitioned into chunks to minimize completion time on parallel systems~\cite{Bharadwaj1996DLT}. While influential in distributed computing, DLT typically assumes offline decomposition and does not address the online, scheduler-interactive mechanisms required in GPU datacenters where workload arrival is dynamic and slice availability fragmented.

(iii) \emph{Runtime tactics: migration, preemption, and elasticity.} 
Recent work on optimal workload placement in MIG environments demonstrates that migration and reallocation can theoretically reduce fragmentation and improve throughput, yet in practice this benefit is often outweighed by the substantial overhead of transferring large GPU states~\cite{Turkkan2024}. Preemption can enhance responsiveness by interrupting ongoing inference tasks and reallocating GPU capacity, but Llumnix shows that such mechanisms incur significant overhead—especially for large language models where state transfers are costly~\cite{Sun2024Llumnix}. Elastic training frameworks such as ElasticFlow~\cite{Gu2023ElasticFlow} enable jobs to dynamically expand or shrink their GPU allocation during runtime (by scaling the number of workers up or down), thereby adapting to cluster load. However, they still rely on checkpointing and resumption when resizing, rather than generating atomized fragments that can be scheduled opportunistically. Beyond these, runtime adaptivity frameworks such as MISO~\cite{Li2022}, which exploits multi-instance GPU capability to support multi-tenant clusters, and MIGRator~\cite{Wang2024MIGRator}, which dynamically reconfigures MIG layouts through job migration, demonstrate that online reconfiguration can mitigate fragmentation and improve utilization. However, both approaches continue to treat jobs as monolithic runs and rely on checkpointing or reallocation, rather than atomizing workloads into restartable subjobs. Systems such as Pollux~\cite{Qiao2021Pollux} exemplify co-adaptive cluster scheduling, where the scheduler dynamically adjusts GPU allocations and jobs respond by tuning internal parameters such as batch size, learning rate, or gradient accumulation. This establishes a genuine bidirectional protocol between jobs and the scheduler, but the adjustments remain intra-job: Pollux modulates training efficiency within a single monolithic run rather than decomposing workloads into restartable subjobs. Recent LLM-oriented systems such as Sarathi-Serve~\cite{Agrawal2024Orion} follow a similar philosophy, dynamically balancing throughput and latency trade-offs via predictive batching strategies. Yet like Pollux, Sarathi-Serve focuses on intra-job tuning rather than cross-job atomization.

(iv) \emph{GPU sharing and isolation.} 
Production systems increasingly depend on mechanisms that allow multiple jobs to share GPU devices. Guardian~\cite{Pavlidakis2024Guardian} addresses security and safety in GPU sharing, while Kubernetes and its device plugin framework provide the foundation for managing fractional GPU allocations~\cite{Kubernetes2024Doc}. These works emphasize isolation and multi-tenancy, but do not propose interactive protocols between scheduler and jobs.

(v) \emph{Cluster-level orchestration.} 
At the datacenter scale, systems such as Google Borg~\cite{Verma2015Borg} and Mesos~\cite{Hindman2011Mesos} established the foundations of multi-tenant cluster management by enabling resource sharing across thousands of nodes. Both incorporate priorities, fairness, and preemptible tasks, but jobs remain indivisible units with no modeling of fine-grained temporal demand. Mesos pioneered the offer-based resource allocation model, where the master advertises available resources and frameworks autonomously decide how to consume them. This mechanism bears a superficial similarity to SJA’s bidirectional protocol, yet Mesos operates at coarse granularity and assumes monolithic tasks—without predictive envelopes or safe-by-construction subjob decomposition. Building on these ideas in the GPU domain, HiveD~\cite{Zhao2020OSDI} introduces deterministic partitioning and elastic sharing policies for deep learning clusters, but like Borg and Mesos, it ultimately treats jobs as monolithic workloads. Recent benchmarking studies such as MIGPerf~\cite{zhang2023migperf} show that fragmentation and interference are inherent to current MIG partitioning schemes and remain bottlenecks even under advanced orchestration frameworks, underscoring the need for fundamentally different scheduling approaches. SJA addresses this by eliminating monolithic placement in favor of subjobs that are safe by construction within advertised gaps.

\subsection{Novelty} 
We introduce SJA, which elevates scheduler–job interaction to a first-class principle: the scheduler announces near-term execution opportunities (time/memory windows on specific slices), and the job responds by generating safe, self-contained subjobs that fit those windows by construction. This bidirectional protocol differs fundamentally from moldable or malleable approaches (which still submit monolithic runs after shape selection), from DLT (which partitions work offline without responding to online slot offers), and from migration or preemption (which react after a job has started). Atomization seeks to achieve higher slice utilization and lower wait times without migration, because placement is correct at admission time; it also integrates naturally with predictive signals such as temporal memory profiles to ensure subjob footprints remain within slice limits. Recent runtime adaptivity frameworks indeed improve utilization through online reconfiguration, yet they continue to operate on monolithic jobs and rely on migration or rescaling. Benchmarking studies such as MIGPerf~\cite{zhang2023migperf} show that fragmentation is inherent to rigid MIG partitioning and persists even under advanced orchestration frameworks. In contrast, atomization avoids mid-run reallocations altogether: subjobs are generated to match the advertised gaps from the outset, thereby mitigating fragmentation without costly state transfers.

Moreover, by decomposing jobs into subjobs that are safe by construction, SJA also reduces the need for complex global optimization in the scheduler. Traditional approaches often rely on sophisticated heuristics or optimization frameworks (e.g., ILP solvers or Gurobi) to balance competing constraints and anticipate future arrivals. Atomization simplifies this task: instead of reasoning far into the scheduling horizon, the scheduler primarily needs to advertise near-term gaps and admit compatible subjobs. This shifts much of the complexity from combinatorial placement into the local generation of subjobs, making the scheduling loop more lightweight and responsive while still improving utilization.

Although SJA shares surface similarity with Mesos in adopting an offer–reply style protocol, the novelty lies in coupling these offers with probabilistic profiles (TRPs/FMPs) and safe-by-construction subjobs. This ensures that admitted work inherently respects slice limits, eliminating the need for reactive reallocation or checkpoint-based recovery.

To the best of our knowledge, no prior work has introduced a scheduler-driven interaction model in which jobs actively generate subjobs in response to \emph{available execution gaps} advertised by the scheduler. Existing approaches such as migration, preemption, or moldable and malleable jobs address flexibility from other angles, but none establish explicit bidirectional communication as the foundation of scheduling decisions.

\section{Concept: Scheduler-Driven Job Atomization}
The core contribution of this paper is the introduction of SJA, a paradigm that redefines how GPU workloads are represented and scheduled. Rather than treating jobs as rigid monolithic entities, atomization conceives them as adaptive and interactive, capable of generating execution fragments that match fragmented resource availability.

To formalize the paradigm, we introduce three key building blocks: (i) the \emph{subjob} as the atomic execution unit, (ii) the \emph{temporal resource profile (TRP)} as a probabilistic description of resource demand over time, and (iii) the \emph{functional memory profile (FMP)} as a statistical summary enabling robust predictions. These concepts establish the foundation of scheduler-driven atomization and will be used consistently throughout the paper.

\subsection{Definition of Subjob}
A \textbf{subjob} is defined as a self-contained execution fragment of a parent job. It is bounded in both \emph{time} (designed to complete within a specific execution window offered by the scheduler) and \emph{resource footprint}. The footprint does not imply constant resource consumption; rather, it denotes that the subjob is confined to a fixed fraction of GPU capacity (e.g., a MIG slice) and that its memory demand is predicted—based on TRP or FMP- to remain below that capacity with high probability throughout its lifetime.

Subjobs may include phases of low usage, warm-up, or bursts, but are only admitted if the scheduler can ensure none of these variations exceed the allocated slice. Unlike preempted jobs, which are interrupted mid-execution, subjobs are deliberately sized to fit their scheduling window and run to completion without external disruption. They may produce intermediate state (e.g., checkpoints or partial outputs), enabling the parent job to resume from the last completed fragment instead of restarting. This design makes subjobs the fundamental building blocks of scheduler-driven atomization.

\subsection{TRP}
A TRP describes how a job’s resource usage evolves over time. Unlike static descriptors such as a single peak estimate, a TRP captures the \emph{temporal structure} of demand, including warm-up phases, steady-state intervals, and bursts. TRPs are probabilistic, reflecting variability across executions of the same job. They may be obtained through dynamic workload profiling, derived from static code analysis, or predicted using models trained on known computational patterns. The essential role of a TRP in scheduler-driven atomization is to provide the scheduler with sufficient information to determine whether a subjob can safely execute within an offered time–capacity window without exceeding resource limits. However, not all workloads naturally lend themselves to stable TRPs. Highly erratic or statistically "spread" resource usage may resist meaningful probabilistic modeling, limiting the scheduler’s ability to guarantee safe placement.

\subsection{FMP}
An \textbf{FMP} is a specialization of the TRP focused on \emph{memory consumption}. It describes how a job’s memory demand evolves over time, including variability across runs, and supports predictions of whether usage will remain below a slice’s capacity throughout a subjob’s lifetime. TRPs and FMPs can be constructed in multiple ways—for example, as pointwise probabilistic bounds or functional summaries such as functional boxplots. The implementation details are flexible; the central requirement is that they capture temporal dynamics of resource usage in a form that supports safe and efficient scheduling decisions.

With these building blocks in place, we now describe the basic principle of scheduler-driven atomization: how schedulers and jobs interact to transform fragmented capacity into safe and efficient execution opportunities.

\begin{figure}[t]
  \centering
  \includegraphics[width=0.9\linewidth]{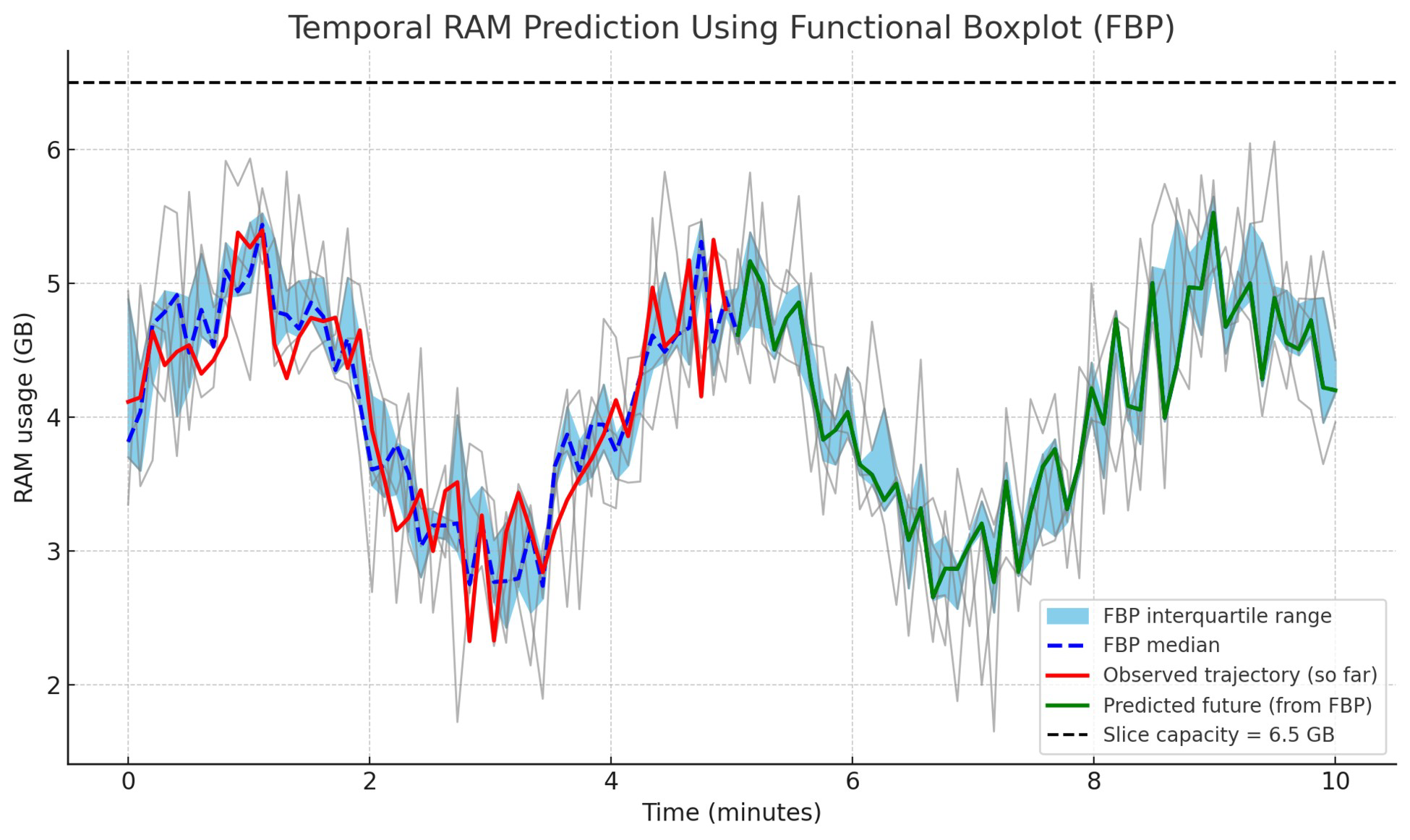}
  \caption{Predicting future RAM usage using a Functional Boxplot (FBP). The red curve shows the observed RAM usage so far. The green curve represents the predicted continuation, based on the median of similar historical trajectories (gray). The blue shaded area indicates the interquartile range (IQR) derived from the FBP. The horizontal dashed line marks the slice capacity. The interquartile range (IQR) indicates the interval between the 25th and 75th percentile of historical RAM values at each time point. It captures the most typical 50\% of outcomes and is used to define the central “band” of the Functional Boxplot.\label{figure_ram_prediction}}
\end{figure}

\subsection{Core Principles}
\label{sec:core-principles}
The design of SJA rests on a set of core principles that distinguish it from existing scheduling paradigms. These principles capture the essence of atomization and explain how it achieves higher utilization, predictability, and extensibility:

\begin{itemize}
    \item \textbf{Jobs as adaptive entities:} jobs are represented not as monolithic blocks but decompose into restartable, self-contained subjobs.
    \item \textbf{Bidirectional communication:} the scheduler advertises execution gaps (time and slice capacity), jobs signal interest if they can safely generate a compatible subjob, and the scheduler applies its policy to decide which job is granted the opportunity. The selected job then materializes the subjob.
    \item \textbf{Safe-by-construction placement:} subjobs are deliberately sized to fit offered windows, avoiding migration, preemption, or mid-run state transfers.
    \item \textbf{Predictive guidance:} TRPs and FMPs provide probabilistic guarantees that subjob footprints remain within slice limits.
    \item \textbf{Fragmentation mitigation:} idle capacity is exploited by packing subjobs into otherwise wasted gaps, substantially reducing fragmentation losses.
    \item \textbf{Reduced scheduling complexity:} the scheduler focuses on advertising near-term windows rather than solving deep global optimizations, while part of the adaptation logic is shifted to the jobs themselves.
    \item \textbf{Policy-agnostic extensibility:} the offer–reply mechanism generalizes naturally to energy-aware, SLO-aware, and fairness-aware scheduling policies.
\end{itemize}

These principles establish the conceptual foundation of SJA. To make them operational, the paradigm requires a concrete mechanism by which schedulers and jobs interact. We therefore now turn to the \emph{protocol and mechanism} of SJA, which specifies how execution opportunities are advertised and how jobs respond with compatible subjobs.

\subsection{Protocol and Mechanism}
Having stated the design principles, we now describe the operational protocol that turns them into practice.

In traditional GPU scheduling, jobs are submitted as rigid monolithic blocks: each declares a peak memory requirement and runtime estimate, and the scheduler allocates an entire slice for the job’s duration. This one-shot allocation often results in poor packing and fragmentation, as slices remain occupied even when the job does not continuously require their full capacity.

SJA redefines scheduling as an \emph{interactive protocol} between the scheduler and jobs. Instead of submitting a fixed block, a job participates in a lightweight negotiation triggered by the scheduler’s offers. The protocol minimally consists of three phases:
\begin{itemize}
\item \textbf{Scheduler $\rightarrow$ Job:} announces available execution windows, defined by time interval and slice capacity.
\item \textbf{Job $\rightarrow$ Scheduler:} signals \emph{interest} if it could safely generate a subjob for that window (but does not materialize it yet), or declines if not applicable. Jobs may also indicate preferences such as deadlines or priority.
\item \textbf{Scheduler:} applies its allocation policy (e.g., fairness, throughput, or SLA constraints) to choose among multiple interested jobs, and grants the slot. Only the selected job then materializes a compatible subjob and execution starts. No subjob state is created until a grant is issued.
\end{itemize}

This minimal three-phase loop forms the core contract of scheduler–job negotiation: the scheduler exposes short-term opportunities, jobs signal potential compatibility, and the scheduler selects which job to grant according to its policy. The chosen job then materializes the subjob. In practice, this exchange can be enriched with additional signals:
\begin{itemize}
\item \emph{Interest:} the job declares that it can fit the offered window and is willing to run if selected.
\item \emph{Decline:} the job indicates that it cannot (or prefers not to) use the offered window.
\item \emph{Preference:} the job attaches constraints (e.g., deadline, checkpoint size, energy budget), which the scheduler may consider when applying its policy.
\end{itemize}

This minimal protocol is deliberately simple—sufficient to enable atomization—yet extensible: additional message types (e.g., priority updates, fairness tokens, or energy signals) can be layered on top without altering the core logic.

Subjobs are deliberately sized to fit their assigned windows and run to completion without interruption. They may produce intermediate state (e.g., checkpoints or partial outputs), enabling the parent job to resume later. This ensures that progress is preserved across multiple fragments without requiring mid-run preemption or migration.

By shifting part of the decision making to the jobs, guided by their TRP, atomization converts scheduling from a static allocation task into a negotiation. The scheduler focuses on advertising near-term gaps, while jobs locally determine whether and how to exploit them. This mechanism directly addresses fragmentation by filling otherwise wasted capacity with safe, restartable work units.

\subsection{Comparison to Existing Paradigms}
\label{sec:comparison}
To clarify the conceptual contribution of SJA, it is helpful to position it against established scheduling paradigms such as moldable jobs, malleable jobs, and migration or preemption. Each of these paradigms represents a different strategy for handling workload diversity and fragmented resources:

\begin{itemize}
    \item \textbf{Moldable jobs} allow the scheduler to decide a resource allocation at submission time, but once chosen, the allocation remains fixed throughout execution.
    \item \textbf{Malleable jobs} provide runtime flexibility by allowing allocations to expand or shrink, but changes are driven entirely by the scheduler and typically require checkpointing.
    \item \textbf{Migration and preemption} mitigate inefficiency by interrupting or relocating jobs after admission, but these techniques often incur significant overhead due to state transfer and disruption.
\end{itemize}

SJA differs fundamentally in that it is \emph{proactive} rather than \emph{reactive}: instead of reshaping or moving already running jobs, it decomposes workloads into restartable subjobs before execution. This design makes fragmentation a first-class scheduling consideration. By relying on lightweight checkpoint-based subjob boundaries, SJA reduces reliance on costly full-state transfers while retaining progress guarantees.

Table~\ref{tab:comparison} summarizes the contrast in the core dimensions of granularity, adaptivity, state movement, and fragmentation handling.

\begin{table}[ht]
\centering
\caption{Comparison of scheduling paradigms with respect to key properties.}
\label{tab:comparison}
{\small
\begin{tabularx}{\textwidth}{p{2cm}XXXX}
\toprule
\textbf{Paradigm} & \textbf{Granularity} & \textbf{Adaptivity} & \textbf{State Move} & \textbf{Fragmentation Handling} \\
\midrule
Moldable & One-time at submission & None at runtime & None & Limited \\
Malleable & Adjustable at runtime & Scheduler-driven & Checkpoint & Partial \\
Migration/ \newline Preemption & Fixed jobs, moved after admission & Scheduler-driven & Full state transfer & Costly \\
SJA & Subjobs tailored to gaps & Bidirectional & Lightweight checkpoint handoff & At admission time \\
\bottomrule
\end{tabularx}
}
\end{table}

\subsection{Relation to Migration}
\label{sec:relation-migration}
A natural question arises: does SJA simply replace one form of migration with another? Traditional \emph{migration} refers to moving a live GPU process across devices or slices while it is running. This approach can reduce fragmentation but incurs substantial overhead, since the entire GPU state (including model weights, optimizer state, and intermediate activations) must be transferred mid-execution. For large deep learning models, such state can reach tens or even hundreds of gigabytes, making mid-run migration prohibitively expensive.

In contrast, SJA deliberately avoids mid-run state transfer. When the scheduler advertises a new execution window on a different slice, a job signals interest but does not move a running process. Only at a \emph{subjob boundary}—a natural cut point where computation can be safely paused—the job materializes a new subjob on the target slice. This involves resuming from a checkpoint or lightweight intermediate state, rather than moving an active GPU context. 

Thus, while both approaches involve execution continuing on a different slice, their mechanics and costs differ fundamentally:
\begin{itemize}
    \item \textbf{Migration:} moves a live process mid-execution, incurring high transfer and coordination overhead.
    \item \textbf{SJA:} re-allocates only at subjob boundaries, where state is compact and progress is preserved without mid-run disruption.
\end{itemize}

By shifting adaptation to subjob granularity, SJA captures the benefits of reallocation without the prohibitive costs of migration. This distinction is essential for making interactive, fine-grained scheduling feasible in practice.

\subsection{Predictive Guidance with TRP}
To ensure subjobs fit into offered gaps, jobs rely on TRPs and, when memory is critical, on FMPs. These probabilistic models describe how resource demand evolves over time and allow the scheduler to estimate the risk of exceeding slice capacity. This predictive layer both filters unsafe subjobs and enables \emph{risk-aware scheduling}: depending on datacenter policy, the scheduler may admit only subjobs with, for example, less than 5\% probability of violation, or relax thresholds to increase utilization under high demand. Figure~\ref{figure_ram_prediction} illustrates this mechanism: the observed trajectory is compared with historical executions, while the functional boxplot provides a probabilistic prediction that ensures memory demand remains within slice capacity throughout the subjob’s lifetime.

Formally, let $M_J(t)$ denote the predicted memory demand of subjob $J$ at time $t$, obtained from its TRP. A subjob $J$ is admissible on a slice of capacity $C$ over an interval $[0,T]$ if
\[
\Pr\!\left( \max_{t \in [0,T]} M_J(t) \leq C \right) \;\geq\; 1 - \epsilon ,
\]
where $\Pr(\cdot)$ is the probability measure induced by the TRP/FMP model and $\epsilon \in (0,1)$ is a configurable risk tolerance (e.g., $\epsilon=0.05$).

\subsection{Offer-Driven Segmentation Policy}
In SJA, segmentation is not performed offline for the entire trajectory. Since the risk-adjusted envelope 
\[
U(t) = Q_{1-\epsilon}[M(t)]
\]
---where $M(t)$ denotes the random memory demand at time $t$ and $Q_{1-\epsilon}$ is the $(1-\epsilon)$ quantile operator---may evolve as runtime predictions are refined, precomputing fixed subjob boundaries would often be wasted effort. Instead, segmentation is applied \emph{on demand}, triggered when the scheduler advertises an execution window. At that moment, jobs evaluate whether they could safely generate a subjob for the offered time/capacity pair and signal interest. Only the selected job, after the scheduler’s policy-based decision, materializes the subjob that fits the allocation.

The guiding principle is \emph{slack-minimizing with hysteresis}: the job maps the offered interval to the \emph{smallest} slice capacity $c_j \in \mathcal{C}$ that
safely covers the predicted demand in that window, i.e.,
\[
\max_{t \in \text{window}} U(t) \;\le\; c_j ,
\]
where $\mathcal{C} = \{c_1, c_2, \dots, c_K\}$ denotes the set of available MIG slice capacities. To avoid pathological over-fragmentation caused by short-lived spikes, two safeguards are enforced: (i) smoothing of $U(t)$ over a short horizon, and (ii) a hysteresis rule that requires a minimum gain in slack before accepting further subdivision. This ensures that segmentation is only applied where it yields meaningful efficiency gains, while otherwise preferring longer, more stable fragments.

Importantly, segmentation policies are not hard-wired: an operator may adjust them according to datacenter priorities. For example, environments prioritizing throughput might allow finer segmentation to pack more subjobs into available gaps, whereas energy-aware or latency-sensitive environments might prefer coarser segmentation to reduce coordination overhead. Leaving policies flexible preserves the generality of SJA as a paradigm, while enabling its instantiations to reflect diverse optimization goals.

\subsection*{Illustrative Scenario}
Consider a GPU cluster where an inference service shares resources with several long-running training jobs. In a traditional scheduler, a training job requesting a 40~GB GPU may wait indefinitely if no full device is available, even though several 10~GB or 20~GB slices remain idle between shorter tasks. Under SJA, the scheduler can advertise a ten-minute execution gap on a 20~GB slice. The training job then generates a compatible subjob that runs safely within this window and produces a checkpoint before yielding resources. Idle capacity is thus filled, the training job progresses instead of stalling, and the inference service retains priority on larger slices. This example illustrates how atomization converts wasted gaps into useful computation while accommodating workload diversity.

\subsection{Benefits}
SJA offers several tangible benefits over conventional scheduling:
\begin{itemize}
    \item \textbf{No migration overhead}: subjobs are sized and placed before execution, avoiding costly checkpointing or mid-run state transfers.
    \item \textbf{Higher utilization}: idle fragments are exploited by packing smaller subjobs into available gaps, reducing fragmentation loss.
    \item \textbf{Reduced wait times}: short subjobs can often start immediately in small gaps, lowering queue latency and improving responsiveness for interactive workloads.
    \item \textbf{Modularity and fault tolerance}: failures or early terminations affect only a single subjob, while the parent job resumes from the last completed fragment.
    \item \textbf{Predictability and risk-awareness}: placement guided by TRPs provides statistically bounded guarantees that subjobs remain within slice capacity throughout execution.
\end{itemize}

\section{Discussion}
\subsection{Comparison to Existing Approaches}
SJA departs from existing scheduling strategies in several important ways. Unlike migration-based techniques, which interrupt a running job and transfer its state elsewhere, atomization generates subjobs that are correctly sized and placed \emph{before} execution. This proactive approach avoids the heavy overheads typically associated with GPU job migration and checkpointing.

Compared with preemption and time-slicing, atomization offers stronger guarantees of isolation and predictability. Preemption may arbitrarily interrupt a job, while time-slicing exposes co-located tasks to interference. Atomized subjobs, by contrast, are self-contained and run to completion within a defined window, reducing the risk of lost progress or unpredictable latency.

The concept also differs from moldable and malleable job models. Moldable jobs allow the scheduler to choose a resource allocation at submission time, while malleable jobs can expand or shrink at runtime. Atomization instead centers on a \textbf{bidirectional protocol}: the scheduler advertises fragmented opportunities, jobs signal interest if they can match them safely, and the scheduler applies its policy to decide which job is granted the slot. Only then does the selected job generate a compatible subjob. This staged interaction distributes decision-making between scheduler and jobs, enabling finer-grained and policy-aware orchestration.

\subsection{Reduction of Scheduling Complexity}
A central benefit of atomization is the simplification of scheduling logic. Conventional approaches, especially when combined with heterogeneous MIG slices, require the scheduler to solve a combinatorial optimization problem, often tackled by heuristics or external solvers such as Gurobi. This global optimization is necessary because jobs are indivisible and must be packed into the schedule as large units.

Atomization mitigates much of this complexity by ensuring that jobs present subjobs already tailored to the scheduler’s advertised windows. The scheduling decision reduces to a lightweight selection problem: from the set of proposed subjobs, the scheduler chooses those that best align with fairness, priority, or service-level objective (SLO) constraints. Importantly, greedy slot-filling is not a limitation in this setting but rather a design choice: subjobs guarantee feasibility at admission time, removing the need for expensive lookahead or reallocation. In this sense, atomization achieves high utilization with dramatically simpler scheduling logic. Optionally, a short lookahead horizon (e.g., tens of minutes) can be used to advertise near-future windows, improving alignment with subjobs without resorting to heavy global optimization.

\subsection{Challenges and Open Issues}
While SJA introduces a fresh perspective on GPU scheduling, it is important to recognize that the paradigm is not without conceptual weaknesses. A fair assessment requires acknowledging the limitations that arise not from implementation hurdles, but from the very assumptions and abstractions on which SJA is built. The following issues highlight critical points where the conceptual foundation of SJA may face tension or fragility.

\begin{enumerate}
    \item \textbf{Abstract nature.}  
    The paradigm assumes that applications can be meaningfully atomized into restartable subjobs. While attractive in theory, it is uncertain whether this assumption holds universally across workloads. Conceptually, this is a weak point: SJA may appear general but may in practice only fit certain classes of applications (e.g., deep learning training or inference).
    
    \emph{Possible mitigation.} We acknowledge that certain workloads are intrinsically hard to atomize. Typical examples include tightly coupled HPC simulations with frequent global synchronization and large, long-lived state (e.g., stencil-based CFD time-steppers or multi-physics solvers), or monolithic GPU kernels whose correctness relies on uninterrupted execution and offer no safe checkpoint boundary. For such cases, SJA should operate in a \emph{hybrid} mode: atomizable jobs follow the offer-reply protocol, whereas non-atomizable jobs fall back to conventional scheduling (e.g., whole-GPU or moldable/malleable placement) or to a more suitable alternative.
    
    \item \textbf{Shift of complexity to jobs.}  
    Simplifying the scheduler sounds promising, but the complexity is implicitly transferred to the application layer. This raises the conceptual question of whether SJA simply exchanges one kind of complexity for another, requiring new APIs, runtime libraries, and changes in frameworks.

    \emph{Possible mitigation.} The apparent transfer of complexity from the scheduler to the job can be alleviated by shifting the burden to \emph{development tools and runtime libraries} rather than to end-users. Instead of requiring application developers to hand-craft atomization logic, frameworks and orchestrators can transparently generate subjob boundaries, insert checkpointing, and expose interfaces to the scheduler. For example, in high-level environments such as R, an orchestrator can automatically instrument user code with the necessary constructs, while in deep learning frameworks atomization could be implemented at the level of training loops and checkpoint managers. In this way, the complexity remains hidden inside libraries and orchestration layers, and application developers continue to program in familiar paradigms without explicit concern for atomization.
    
    \item \textbf{Protocol versus global optimality.}  
    By restricting the scheduler to advertising only short-term slots, SJA avoids heavy optimization overhead. At the same time, it disregards long-term global optimality. Conceptually, this is a clear trade-off: simplicity and local efficiency versus potential loss at the macro scale.

    \emph{Possible mitigation.}
    By mapping each TRP/FMP segment to the lowest slice capacity that safely covers it, the scheduler achieves near-optimal local packing without global search. While global optimality is not guaranteed, this policy provides a tractable middle ground: local efficiency is high, overhead is low, and the complexity of heavy optimization frameworks is avoided. For scenarios where local policies consistently underperform (e.g., under extreme contention), hybrid schedulers could occasionally trigger global re-optimization, combining the strengths of both worlds.
    
    \item \textbf{Dependence on TRP/FMP quality.}  
    The paradigm assumes the existence of reliable predictive profiles. But if TRPs or FMPs are inaccurate or unsuitable, SJA could degenerate into pseudo-random allocations. This shows that SJA is not a self-contained concept—it critically depends on the predictive layer beneath it.

    \emph{Possible mitigation.} This issue can be addressed on several levels. First, workloads with excessively scattered or highly variable TRPs can be filtered out and scheduled conservatively, as atomization is not suitable in such cases. Second, predictions can be corrected dynamically by incorporating online measurements of memory usage and runtime characteristics, turning TRPs into adaptive profiles. Third, confidence thresholds need not be static; schedulers can adjust risk tolerance adaptively according to current cluster load. Fourth, multiple predictors or domain-specific heuristics may be combined, for example leveraging model and batch size in DL training or algorithmic structure in HPC. Finally, predictors may incorporate job input data, enabling resource forecasts that reflect not only past runs but also dataset size or hyperparameters.
    
    \item \textbf{Granularity trap.}  
    Subjobs that are too small may introduce coordination overheads greater than fragmentation losses, while overly large subjobs may fail to reduce fragmentation effectively. Conceptually, SJA must balance carefully on this fine edge, which is non-trivial.
    
    \emph{Possible mitigation.}
    Instead of fixed-size subjobs, workloads are segmented according to TRP/FMP structure: each fragment is chosen to fit the smallest feasible slice without exceeding it, which naturally avoids both overly small and excessively large subjobs. Jobs with smooth or slowly varying demand produce longer subjobs, while bursty workloads result in finer segmentation only where necessary, preventing unnecessary overhead. Minimum and maximum subjob duration thresholds can be enforced at the framework level, ensuring that coordination costs remain bounded and that subjobs retain useful granularity.    
\end{enumerate}

Taken together, these challenges do not invalidate the paradigm but underscore that SJA is best viewed as a research direction rather than a ready-to-deploy solution. Its conceptual promise lies in shifting the locus of scheduling from monolithic allocation toward interactive job–scheduler negotiation. Even if only certain workload classes (e.g., deep learning training or inference) ultimately benefit, demonstrating viability in these domains would already mark a significant advance and justify further exploration of the paradigm.

\subsection{Future Opportunities}
Beyond immediate efficiency gains, SJA’s protocol-centric design enables specializations that address fundamental system goals. We outline several directions where extending the offer$\rightarrow$reply schema yields qualitatively new capabilities:

\begin{enumerate}
    \item \textbf{Energy- and carbon-aware atomization.}
    Energy efficiency is a critical concern in GPU datacenters, where idle slices still consume significant static power. By packing opportunistic subjobs into otherwise unused capacity, atomization increases effective performance-per-watt and reduces wasted consumption. Moreover, the protocol allows offers to carry information about power caps, dynamic electricity prices, or grid carbon intensity, while subjobs respond with predicted energy traces ($\widehat{P}(t)$, $\widehat{E}$). Admission control can then enforce power and risk constraints, for example
    \[
      \Pr\!\big(\max_{t\in[0,T]} P(t)\le \text{power\_cap}\big)\ge 1-\alpha_P,
    \]
    and shift flexible subjobs into low-carbon or low-tariff windows. This makes atomization a natural building block for sustainable HPC and cloud operations.

    \item \textbf{Deadline- and SLO-aware scheduling.}
    Offers carry timing and priority; subjobs return predictive runtimes and deadlines. An Earliest Deadline First (EDF)-like policy on windows admits only subjobs whose completion time $T$ (a random variable for runtime) is below their deadline $\tau$ with sufficiently high probability: 
    \[
    \Pr(T \leq \tau) \geq 1 - \alpha_T,
    \]
    where $\tau$ is the required deadline or SLO bound and $\alpha_T$ is the tolerated risk of violation (e.g., $\alpha_T = 0.05$). This ensures that atomized fragments align with latency or throughput SLOs.

    \item \textbf{Fairness with budgets or tokens.}
    Tenants receive credits that are consumed by admitted subjobs. The scheduler monitors slowdown or Jain’s index and adjusts offers to maintain fairness while exploiting fragmented capacity.

    \item \textbf{Market mechanisms.}
    Windows can be priced; subjobs bid with a value function subject to risk and energy constraints. This supports welfare-maximizing allocation and exposes fragmentation as a tradable resource.

    \item \textbf{State- and locality-aware placement.}
    Subjobs declare predicted checkpoint size $\widehat{S}$ and locality preferences (e.g., same slice for warm caches). Admission penalizes large $\widehat{S}$ and favors locality, reducing orchestration overheads.

    \item \textbf{Heterogeneous accelerators.}
    The same protocol generalizes beyond GPUs to CPUs, VPUs, or FPGAs. Offers expose per-device windows and capacities; subjobs choose feasible targets via their TRPs (with FMP as the memory-specialized instance).
\end{enumerate}

\section{Future Work}
We plan a trace-driven simulation on MIG-like clusters comparing SJA against baselines: (i) best-fit/first-fit packing with monolithic jobs, (ii) moldable/malleable scheduling, and (iii) migration/preemption where available. Metrics: GPU RAM utilization, queueing delay, rejection rate, number of interruptions, and performance per watt (where power traces are available). Ablations will study the impact of risk tolerance $\epsilon$, minimum subjob duration $\tau_{\min}$, smoothing windows for $U(t)$, short-horizon lookahead, and predictor precision (offline only vs. offline + online correction). Workloads will include DL training/inference (with varied models and batch sizes), synthetic bursty profiles, and selected HPC kernels (including non-atomizable cases as negative controls).

While this paper introduces the concept of SJA, much remains to be explored before the paradigm can be fully realized and deployed in production systems. We highlight several promising directions for future research.

\subsection{Simulation-Based Validation}
The first step will be to develop large-scale simulation frameworks that quantify the benefits of atomization under realistic workload conditions. These simulations will compare atomization with conventional heuristics (e.g., First-Fit, Best-Fit), moldable or malleable scheduling, and migration-based strategies. Metrics such as GPU utilization, wait-time distribution, job rejection rates, and wasted compute will be evaluated to provide statistical evidence of the paradigm's advantages.

\subsection{Integration with Predictive Models}
The effectiveness of atomization depends on accurate TRPs. Future work will refine these models by combining offline profiling with online observations and by applying machine learning techniques to improve prediction accuracy. This includes exploring probabilistic thresholds that balance safety with utilization.

\subsection{Reinforcement Learning for Subjob Generation}
Atomization enables policy-driven generation of subjobs. A promising direction is to apply Reinforcement Learning (RL) to guide decisions about when to accept, reject, or reshape subjobs. Such policies could optimize multiple objectives simultaneously, including utilization, fairness, and energy efficiency.

\subsection{Fairness and Multi-Tenant Policies}
In multi-tenant datacenters, fairness is as important as efficiency. Future work will investigate how atomization can incorporate fairness constraints, ensuring that no user or workload type is systematically disadvantaged by the scheduler's decisions.

\subsection{Extension Beyond GPUs}
Although this paper focuses on GPU workloads and MIG-enabled architectures, the concept of scheduler-driven atomization has broader applicability. Future investigations will extend the paradigm to CPU scheduling, hybrid CPU-GPU clusters, and containerized cloud environments where fragmented resources frequently arise.

\subsection{System Prototyping}
Practical deployment will require prototyping in real GPU clusters. This involves integrating atomization into existing orchestration frameworks such as Kubernetes~\cite{Kubernetes2024Doc}, evaluating runtime overhead, and developing interfaces that allow jobs to interact with the scheduler through a standardized protocol.

\noindent
As an initial step, we plan to build a prototype within the \textbf{R programming ecosystem}, exposing the scheduler as an R package for reproducible experiments, with \textbf{CUDA} as the computational engine to support GPU-accelerated workloads and to demonstrate feasibility in practice.

\subsection{Integration with Kubernetes}
A natural path forward is to integrate scheduler-driven atomization into existing container orchestration frameworks, most notably Kubernetes. Kubernetes already provides core primitives for cluster management, including pods, resource requests, and extensible scheduling via plugins. GPU devices are exposed through vendor-specific device plugins (e.g., the NVIDIA device plugin), and with the advent of MIG, each physical GPU can be partitioned into several isolated slices that Kubernetes can allocate individually.

Today, however, Kubernetes scheduling remains largely static: a pod either receives an entire GPU or a statically configured MIG slice. If the requested resource is not available, the job simply waits. The system neither negotiates fragmented time slots nor reconfigures MIG on demand, and it does not enable jobs to generate subjobs in response to real-time opportunities.

Scheduler-driven atomization could be implemented as a \emph{custom scheduling plugin} for Kubernetes. In this design, Kubernetes provides the orchestration substrate, while atomization logic governs how fragmented GPU capacity is offered to jobs and how jobs respond with compatible subjobs. Prototyping this approach in tandem with the R/CUDA stack described above will enable reproducible experimentation and demonstrate feasibility in Kubernetes-based environments.

\subsection{Beyond Prediction: Toward Rich Scheduler-Application Communication}
While this paper has focused on temporal prediction and atomization, we believe that the long-term trajectory of GPU scheduling will extend far beyond prediction alone. The deeper trend is that applications will no longer remain passive black boxes but will increasingly \emph{communicate} with the infrastructure. This shift is motivated by three converging forces: (i) hardware capabilities such as MIG, power-state tuning, and mixed-precision modes that can only be exploited if runtime behavior is visible to the scheduler; (ii) software ecosystems such as PyTorch/XLA and TensorFlow runtimes, which already provide hooks for performance monitoring and adaptation; and (iii) economic pressure to maximize performance-per-watt in large GPU clusters.

In such a future, applications may not only expose TRPs but also provide richer signals: upcoming critical phases, tolerance to reduced precision, checkpoint sizes, or energy/fairness preferences. Conversely, schedulers may embed policies that reflect energy price fluctuations, carbon intensity, or service-level objectives (SLOs). The resulting dialog transcends prediction: it forms a continuous feedback loop in which jobs adapt their structure, precision, and parallelism in response to scheduling opportunities, while the scheduler orchestrates resources with global efficiency and fairness in mind.

We see SJA as a natural first step toward this broader vision. Atomization establishes the protocol of offer $\rightarrow$ reply, proving that jobs can adaptively reshape themselves when given suitable information. Extending this principle beyond prediction-based placement to energy-, accuracy-, or QoS-aware negotiation represents, in our view, the most promising and conceptually coherent path for the evolution of GPU scheduling.

\section{Conclusion}
This concept paper introduced SJA, a paradigm in which schedulers and jobs interact through a bidirectional protocol to transform fragmented GPU capacity into safe execution opportunities. Unlike migration, preemption, or moldable models, SJA decomposes workloads into subjobs guided by probabilistic resource profiles, ensuring that each fragment fits its allocated slice by construction.

The paradigm differs from existing strategies by making scheduler–job communication explicit, shifting part of the decision-making to jobs themselves, and reducing reliance on costly mid-run interventions. It opens a new path toward higher utilization, lower wait times, and risk-aware scheduling in MIG-enabled datacenters.

As a concept paper, this work deliberately emphasized principles over experiments. Future research will provide simulation-based validation, refined predictive models, policy learning for subjob generation, and practical prototyping in the R/CUDA stack and Kubernetes environments. By moving from static jobs to adaptive subjobs, scheduler-driven atomization points toward a next generation of efficient and predictable workload management. By design, SJA prioritizes local feasibility and responsiveness over global optimality; in our view this is a principled trade-off that unlocks substantial efficiency gains without heavy planning overhead.

\bibliographystyle{plainurl}

\end{document}